\documentclass[pra,aps,reprint,showpacs,amsmath]{revtex4-1}

\usepackage{bm}
\usepackage{graphicx}
\usepackage{dsfont}

\newcommand{\bra}[1]{\langle #1 | \,}
\newcommand{\ket}[1]{\, | #1 \rangle}
\newcommand{\braket}[2]{\langle #1 | #2 \rangle}
\newcommand{\expv}[1]{\langle #1 \rangle}

\newcommand{\Om}{\Omega}
\newcommand{\ga}{\gamma}
\newcommand{\Ga}{\Gamma}

\newcommand{\la}{\lambda}

\newcommand{\hlf}{\frac{1}{2}}
\newcommand{\mc}[1]{\mathcal{#1}}
\newcommand{\sig}{\hat{\sigma}}
\newcommand{\Sig}{\hat{\Sigma}}
\newcommand{\hrho}{\hat{\rho}}

\begin{document}

\title{Stimulated adiabatic passage in a dissipative Rydberg superatom}
 
\author{David Petrosyan}
\affiliation{Institute of Electronic Structure and Laser, FORTH, 
GR-71110 Heraklion, Crete, Greece}

\author{Klaus M\o lmer}
\affiliation{Department of Physics and Astronomy, University of Aarhus,
DK-8000 Aarhus C, Denmark}

\date{\today}

\begin{abstract}
We study two-photon excitation of Rydberg states of atoms 
under stimulated adiabatic passage with delayed laser pulses. 
We find that the combination of strong interaction between the atoms 
in Rydberg state and the spontaneous decay of the intermediate 
exited atomic state leads to the Rydberg excitation of precisely 
one atom within the atomic ensemble. The quantum Zeno effect 
offers a lucid interpretation of this result: 
the Rydberg blocked atoms repetitively scattering photons
effectively monitor a randomly excited atom which therefore 
remains in the Rydberg state. This system can be used for 
deterministic creation and, possibly, extraction of Rydberg atoms 
or ions one at a time. The sympathetic monitoring via decay of 
ancilla particles may find wider applications for
state preparation and probing of interactions in 
dissipative many-body systems.
\end{abstract}


\maketitle

\section{Introduction}

Rydberg atoms strongly interact with each other via long-range 
dipole-dipole (DD) or van der Waals (vdW) potentials \cite{RydAtoms}. 
Within a certain interatomic distance, the interaction-induced level 
shifts can suppress resonant optical excitation of multiple Rydberg atoms
\cite{Jaksch2000,Lukin2001,Vogt2006,Tong2004,Singer2004,Heidemann2007,Ates2007,UrbanGaetan2009}. 
A collection of atoms in the corresponding blockade volume then forms 
a ``superatom'' which can accommodate at most one shared Rydberg excitation 
\cite{Robicheaux2005,Stanojevic,Honer2011,Dudin2012NatPh}. 

The Rydberg blockade mechanism constitutes the basis for a number 
of promising quantum information schemes \cite{Jaksch2000,Lukin2001,rydrev} 
and interesting multiatom effects 
\cite{Weimer2008,Low2009,Mueller2009,Saffman2009,Honer2010,Weimer10,Pupillo2010,Schachenmayer2010,Pohl2010,Sela2011,Schwarzkopf2011,Bijnen2011,Viteau2011,Lesanovsky2011,Ji2011,Lesanovsky2012,Ates2012,Schauss2012,Petrosyan2012,Hoening2013}. 
Resonant two-photon excitation of Rydberg states is employed in several schemes 
\cite{Hoening2013,Friedler2005,schempp10,adams10,pohl11,Petrosyan2011,Gorshkov2011,Peyronel2012,RdSV2013,Guenter2012,Pritchard12,Dudin2012Sci,Moeller2008} 
utilizing the effects of atomic coherence, such as electromagnetically 
induced transparency \cite{EIT} and coherent population trapping and 
transfer \cite{stirap}. Stimulated adiabatic passage with delayed pulses
in an ensemble of three-level atoms was previously considered in 
\cite{Moeller2008}, where all the atomic states were assumed to be stable, 
while the lower atomic transition was driven either by a microwave field 
or by  a pair of optical fields in the Raman configuration. It was shown 
that, under the Rydberg blockade, the application of the ``counterintuitive'' 
pulse sequence results in a multiatom entangled state with strongly 
correlated population of the two lower states.

The purpose of the present work is to investigate the more typical experimental 
situation \cite{schempp10,adams10,Peyronel2012,Dudin2012Sci,RdSV2013} 
in which both transitions of three-level atoms 
are driven by optical fields in a ladder ($\Xi$) configuration, while 
the intermediate excited state of the atoms undergoes rapid spontaneous 
decay. For non-interacting (distant) atoms, the situation is analogous 
to what is usually referred to as stimulated Raman adiabatic passage 
(STIRAP) in a $\Lambda$-configuration \cite{stirap}. 
Adding interatomic interactions leads to highly non-trivial behavior
of the Rydberg superatom. In the earlier part of the process,
the system is in a completely symmetric superposition of $N$ atoms, 
each undergoing adiabatic passage towards the Rydberg state without 
populating the intermediate excited state. But once any one atom is
excited to the Rydberg state, it blocks further Rydberg excitations 
and triggers the cycling excitation and decay of the intermediate excited 
state of all the other $N-1$ atoms. This destroys the interatomic coherences 
and dephases the single Rydberg excitation which therefore decouples 
from the field. Through the exact solution of the $N$-atom master equation, 
we obtain at the end of the process a mixed state of the system with 
a single Rydberg excitation incoherently shared among all $N$ atoms. 
We can also understand the underlying physical mechanism in terms of
the quantum Zeno effect \cite{Zeno}, in which atoms emitting spontaneous 
photons through the decay of the intermediate excited state reveal that 
interactions block their adiabatic passage towards the Rydberg state 
and thereby perform frequent projective measurements of the presence 
of a Rydberg excitation in the ensemble.
Remarkably, the larger is the number of atoms within the blockade volume 
the more robust is the transfer process resulting in a single Rydberg 
excitation of the superatom.

\section{Adiabatic passage in a multiatom system}

\subsection{Single-atom STIRAP}

\begin{figure}[t]
\includegraphics[width=8.5cm]{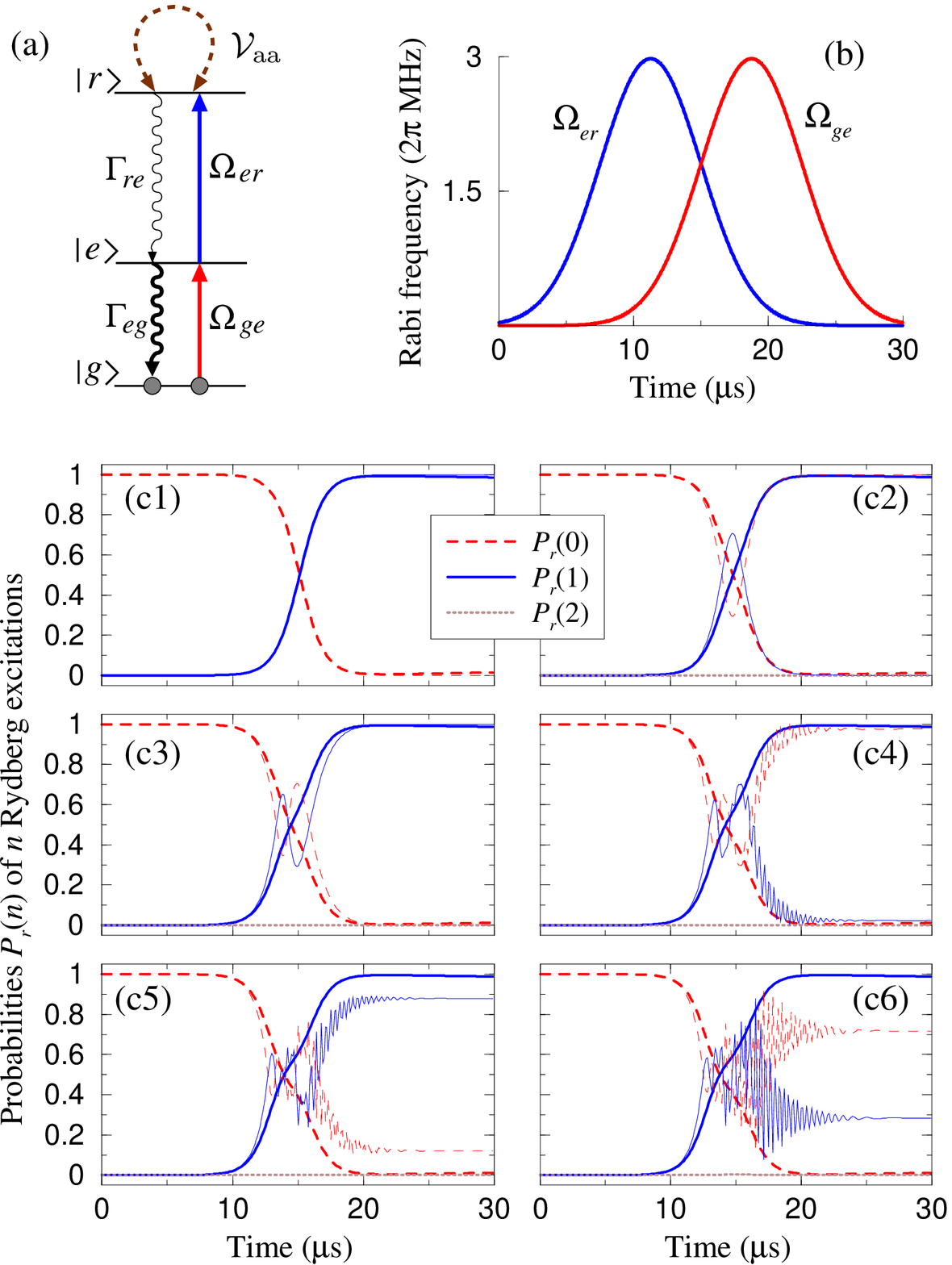}
\caption{
(a) Level scheme of atoms interacting with the fields $\Om_{ge}$ and 
$\Om_{er}$ on the transitions $\ket{g} \to \ket{e}$ and $\ket{e} \to \ket{r}$,
while $\Ga_{eg}$ and $\Ga_{re}$ are (population) decay rates of states 
$\ket{e}$ and $\ket{r}$. $\mc{V}_{\mathrm{aa}}$ denotes the interaction 
between atoms in Rydberg state $\ket{r}$.
(b)~Time dependence of the $\Om_{ge}$ and $\Om_{er}$ fields.
(c)~The corresponding time-dependent probabilities $P_r(n)$ of 
$n=0,1,2$ Rydberg excitations in a compact ensemble of $N = 1-6$ atoms, 
(c1)-(c6), obtained from the exact solutions of Eq.~(\ref{rhoME});
thicker lines correspond to the dissipative system 
($\Ga_{eg},\Ga_{re}$ given in the text), 
while thinner lines to the fully coherent dynamics of 
non-decaying atoms ($\Ga_{eg},\Ga_{re} \simeq 0$).} 
\label{fig:alsOmPr}
\end{figure}

Let us first recall the essence of adiabatic transfer of population 
in an isolated three-level atom using a pair of delayed laser pulses 
(STIRAP) \cite{stirap}.
A coherent optical field with Rabi frequency $\Om_{ge}$ resonantly couples
the stable ground state $\ket{g}$ to an unstable (decaying) excited state
$\ket{e}$, which in turn is resonantly coupled to another stable state 
$\ket{r}$ by the second coherent field of Rabi frequency $\Om_{er}$ 
[Fig.~\ref{fig:alsOmPr}(a)].
The eigenstates of the corresponding Hamiltonian 
$\mc{V}_{\mathrm{af}} = \hbar(\Om_{ge} \ket{e}\bra{g} 
+ \Om_{er} \ket{r}\bra{e} + \mathrm{H. c.})$ are given by
$\ket{\psi_0} = (\cos \theta \ket{g} - \sin \theta \ket{r}$)
and $\ket{\psi_{\pm}} = \frac{1}{\sqrt{2}}
(\sin \theta \ket{g} \pm \ket{e} + \cos \theta \ket{r})$, where 
the mixing angle $\theta$ is defined via $\tan \theta = \Om_{ge}/\Om_{er}$.
The ``dark'' state $\ket{\psi_0}$ with energy $\la_0 = 0$ 
does not have any contribution from the fast decaying state $\ket{e}$, 
while the ``bright'' states $\ket{\psi_{\pm}}$ having energies 
$\la_{\pm} = \hbar \sqrt{\Om_{ge}^2 + \Om_{er}^2}$ do contain $\ket{e}$
and thus are unstable against spontaneous decay. 
The aim of the STIRAP process is to completely transfer the population 
between the two stable states $\ket{g}$ and $\ket{r}$ without  
populating the unstable state $\ket{e}$, which is achieved by
adiabatically changing the dark state superposition.
With the system initially in state $\ket{g}$, one first applies
the $\Om_{er}$ field, resulting in $\braket{g}{\psi_0} = 1$
($\Om_{ge} \ll \Om_{er}$ and therefore $\theta = 0$).
This is then followed by switching on $\Om_{ge}$ and 
switching off $\Om_{er}$ [Fig.~\ref{fig:alsOmPr}(b)], 
resulting in $|\braket{r}{\psi_0}| = 1$
($\Om_{ge} \gg \Om_{er}$ and therefore $\theta = \pi/2$). 
If the mixing angle is rotated slowly enough, 
$\dot{\theta} \ll \frac{1}{\hbar} |\la_{\pm} - \la_0|$,
the system adiabatically follows the dark state $\ket{\psi_0}$, and
the bright states $\ket{\psi_{\pm}}$, and thereby $\ket{e}$, are 
never populated. Hence, the decay of $\ket{e}$ is neutralized and 
the population of the system is completely transferred from $\ket{g}$
to $\ket{r}$. For what follows, it is useful to remember that
the dark state $\ket{\psi_0}$ does not contain $\ket{e}$ because the
resonant coupling of $\ket{g}$ to $\ket{e}$ by $\Om_{ge}$ interferes 
destructively with the resonant coupling of $\ket{r}$ to $\ket{e}$ 
by $\Om_{er}$.

\subsection{The $N$-atom master equation}

Consider now an ensemble of $N$ three-level atoms confined in a small 
volume with linear dimension $L$ of several $\mu$m. All the atoms 
uniformly interact with two optical fields of Rabi frequencies 
$\Om_{ge}$ and $\Om_{er}$ as shown in Fig.~\ref{fig:alsOmPr}(a). 
The atom-field interaction Hamiltonian reads $\mc{V}^j_{\mathrm{af}} 
= \hbar(\Om_{ge} \sig_{eg}^j + \Om_{er} \sig_{re}^j + \mathrm{H. c.})$, 
where $\sig_{\mu \nu}^j \equiv \ket{\mu}_{jj}\bra{\nu}$ 
are the transition operators for atom $j$. The intermediate excited state 
$\ket{e}$ decays to the ground state $\ket{g}$ with the rate $\Ga_{eg}$; 
the corresponding Liouvillian acting on the density matrix $\hrho$ of the 
system is given by $\mc{L}_{eg}^j \hrho = \hlf \Ga_{eg} 
[2 \sig_{ge}^j \hrho \sig_{eg}^j - \sig_{ee}^j \hrho - \hrho \sig_{ee}^j]$.
The decay rate $\Ga_{re}$ of the highly excited Rydberg state $\ket{r}$ 
is typically much smaller 
(and can be neglected when $\Ga_{re} \ll \Om_{er}^2/\Ga_{eg}$),
but for completeness we include it via $\mc{L}_{re}^j \hrho = \hlf \Ga_{re} 
[2 \sig_{er}^j \hrho \sig_{re}^j - \sig_{rr}^j \hrho - \hrho \sig_{rr}^j]$.
Note that both transitions of the three-level atoms are assumed closed.

We next include the interatomic interactions. 
The long-range potential between pairs of atoms $i,j$ 
in the Rydberg state $\ket{r}$ induces level shifts 
$\Delta_{ij} = C_p/d_{ij}^{p}$ of states $\ket{r_i r_j}$, where 
$d_{ij}$ is the interatomic distance and $C_p$ is the DD ($p=3$) 
or vdW ($p=6$) coefficient. The atom-atom interaction Hamiltonian reads
$\mc{V}_{\mathrm{aa}}^{ij} = \hbar  \sig_{rr}^i \Delta_{ij} \sig_{rr}^j$.
We assume that all the atoms are within a blockade distance from each other, 
$\Delta_{ij} \gg \max[w] \; \forall \; i,j \in [1,N]$, where 
$w = \frac{\Omega_{ge}^2 +\Omega_{er}^2 }{\sqrt{2 \Omega_{ge}^2 + \Ga_{eg}^2/4}}$
is the Rydberg-state excitation linewidth of a single three-level atom. 
We define the probabilities $P_r(n) = \expv{\Sig_r^{(n)}}$ of $n$ Rydberg 
excitations of superatom through the corresponding projectors 
$\Sig_{r}^{(0)} \equiv \prod_{i=1}^{N} (\sig_{gg}^i +\sig_{ee}^i) 
= \prod_{i=1}^{N} (\mathds{1} - \sig_{rr}^i)$,  
$\Sig_{r}^{(1)} \equiv \sum_{j=1}^{N} \sig_{rr}^j
\prod_{i \neq j}^{N} (\mathds{1} - \sig_{rr}^i)$, etc.
Note that $\sig_{gg}^i +\sig_{ee}^i + \sig_{rr}^i = \mathds{1} \; 
\forall \; i \in [1,N]$.  

The density operator $\hrho$ of the $N$-atom system obeys the master equation
\cite{PLDP2006} 
\begin{equation}
\partial_t \hrho = -\frac{i}{\hbar} [\mc{H}, \hrho] +  \mc{L} \hrho , 
\label{rhoME}
\end{equation}
with the Hamiltonian 
$\mc{H} = \sum_j \mc{V}_{\mathrm{af}}^j + \sum_{i<j} \mc{V}_{\mathrm{aa}}^{ij}$ 
and the Liouvillian 
$\mc{L} \hrho =  \sum_j (\mc{L}_{eg}^j \hrho + \mc{L}_{re}^j \hrho)$.

\section{Numerical simulations}

We solve the master equation~(\ref{rhoME}) numerically assuming the 
atoms are irradiated by the two pulsed fields having Gaussian temporal shapes 
\[
\Om_{ge,er}(t) = \Om_0 \exp \left[ - \frac{( t - \hlf t_{\mathrm{end}} 
\mp \sigma_t )^2} {2 \sigma_t^2} \right] ,
\] 
where $\Om_0 = 2 \pi \times 3\:$MHz is the peak amplitude, 
$2\sigma_t =\frac{1}{4} t_{\mathrm{end}}$ is the temporal width 
and relative delay of the pulses, and $t_{\mathrm{end}} = 30\:\mu$s 
is the process duration [see Fig.~\ref{fig:alsOmPr}(b)]. We take cold 
$^{87}$Rb atoms \cite{schempp10,adams10,Peyronel2012,Dudin2012Sci,RdSV2013}, 
with the ground state 
$\ket{g} \equiv 5 S_{1/2} \ket{F=2,m_F=2}$, the intermediate excited state 
$\ket{e} \equiv 5 P_{3/2} \ket{F=3,m_F=3}$ with 
$\Ga_{eg} = 38\:$MHz, and the highly excited Rydberg 
state $\ket{r} \equiv n S_{1/2}$ with the principal quantum number $n \sim 80$
and $\Ga_{re} = 1\:$KHz. Within the trapping volume of linear dimension 
$L \sim 5\:\mu$m we then have large interatomic (vdW) interactions 
\cite{rydcalc} $\Delta_{ij} \geq 10 w_0 \; \forall \; d_{ij} \leq L$, 
where $w_0 = \frac{2 \Omega_{0}^2}{\sqrt{2 \Omega_{0}^2 + \Ga_{eg}^2/4}} 
\simeq 2 \pi \times 3.5\:$MHz. 

The results of simulations for $N=1,\ldots,6$ atoms are summarized in 
Fig.~\ref{fig:alsOmPr}(c). For any $N$, even or odd, the ``counterintuitive''
sequence of pulses $\Om_{ge,er}(t)$ leads, with large probability 
$P_r(1) \geq 0.98$, to a single Rydberg excitation of the superatom, while 
the probabilities of multiple excitations $P_r(n > 1)$ are negligible,
due to the strong blockade. Once a Rydberg excitation is produced, the 
small decay $\Ga_{re}$ of state $\ket{r}$ leads to a slow decrease of $P_r(1)$.

The response of the Rydberg superatom to the ``counterintuitive'' 
sequence of pulses may look analogous to the coherent adiabatic passage 
of a single three-level atom, but this similarity is superficial 
and the physics behind it is more involved.
This is perhaps best illustrated in Fig.~\ref{fig:alsOmPr}(c) 
by the strikingly different behavior of superatom in the absence of 
dissipation, $\Ga_{eg},\Ga_{re} = 0$, which was studied in \cite{Moeller2008}.
Without dissipation, in the transition region  
$\Om_{ge}(t) \sim \Om_{er}(t)$ the probabilities of zero $P_r(0)$ and 
one $P_r(1)$ Rydberg excitation do not change monotonically but 
alternate $N-1$ times, with the result that for an even number of atoms $N$ 
the final state of the system does not contain a Rydberg excitation. 
[For $N \geq 4$, the fast oscillations of probabilities $P_r(0,1)$ 
and their final values noticeably different from $0$ and $1$ 
are due to the violation of adiabaticity with the increased 
system size and the corresponding decrease in the separation 
between its eigenstates \cite{Moeller2008}].

\subsection{Analysis}

For a dissipationless system, it is convenient to use the fully 
symmetrized states $\ket{n_g, n_e, n_r}$ denoting $n_g$ atoms 
in state $\ket{g}$, $n_e$ atoms in $\ket{e}$, and $n_r$ atoms 
in $\ket{r}$. Due to the Rydberg blockade, only $n_r = 0,1$ values are 
allowed, while $n_g + n_e + n_r = N$. The field $\Om_{ge}$ couples the 
ground state of the superatom $\ket{N_g, 0_e, 0_r}$ successively to the 
collective single $\ket{(N-1)_g, 1_e, 0_r}$, double $\ket{(N-2)_g, 2_e, 0_r}$ 
etc. excitation states, which are in turn coupled to the single Rydberg 
excitation states $\ket{(N-1)_g, 0_e, 1_r}$, $\ket{(N-2)_g, 1_e, 1_r}$ etc. 
by the field $\Om_{er}$ \cite{Petrosyan2011}. The corresponding Hamiltonian 
can be expressed as $\mc{H} = \hbar (\Om_{ge} \hat{e}^{\dag} \hat{g} + 
\Om_{er} \hat{r}^{\dag} \hat{e}+ \mathrm{H. c.})$, where
operators $\hat{g}$ ($\hat{g}^{\dag}$), $\hat{e}$ ($\hat{e}^{\dag}$)
and $\hat{r}$ ($\hat{r}^{\dag}$) annihilate (create) an atom in the 
corresponding state $\ket{g}$, $\ket{e}$ and $\ket{r}$; 
$\hat{g}$ and $\hat{e}$ are standard bosonic operators, while
$\hat{r}$ describes a hard-core boson $(\hat{r}^{\dag})^2 =0$.
As was shown in \cite{Moeller2008} for non-decaying atoms, 
an ideal adiabatic passage leads to the final state of the system 
$\ket{J_x = 0}$ for $N$ even, and $\ket{J_x = 0} \otimes \ket{1_r}$ for $N$ odd,
where $\ket{J_x = 0}$ is the eigenstate of operator
$\hat{J}_x \equiv \frac{1}{2} (\hat{e}^{\dag} \hat{g} + \hat{g}^{\dag} \hat{e})$
with zero eigenvalue. The state $\ket{J_x = 0}$ involves an equal number 
of atoms [$N/2$ or $(N-1)/2$] in  $(\ket{g} \pm \ket{e})/\sqrt{2}$. 

In the presence of strong decay $\Ga_{eg} \gtrsim \Om_{ge}$ of 
state $\ket{e}$, the dynamic of the system is completely different. 
If we had non-interacting atoms, the adiabatic passage would yield a 
product state $(\cos \theta \ket{g} - \sin \theta \ket{r})^{\otimes N}$
containing multiple Rydberg excitations but no atoms in state $\ket{e}$. 
The strong interatomic interactions, however, shift the energies of multiply 
excited Rydberg states out of resonance with the $\Om_{er}$ field. 
Starting from the ground state $\ket{N_g, 0_e, 0_r} = \prod_{i=1}^{N} \ket{g}_i$ 
the population transfer beyond the symmetric 
single Rydberg excitation state $\ket{(N-1)_g, 0_e, 1_r} = 
\frac{1}{\sqrt{N}} \sum_{j=1}^{N} \ket{r}_j \prod_{i \neq j}^{N} \ket{g}_i$
is then blocked. In this superposition, the atoms in state $\ket{g}$ can  
be now excited to state $\ket{e}$ by the strong resonant field $\Om_{ge}$ 
since the coupling of $\ket{e}$ to $\ket{r}$ by $\Om_{er}$ and the 
resulting destructive interference are suppressed. Atoms excited 
to $\ket{e}$ rapidly decay back to $\ket{g}$ with random phases. 
This leads to continuous dephasing of the superposition
$\ket{(N-1)_g, 0_e, 1_r}$, turning it into an incoherent mixture of single
Rydberg excitation, which is decoupled from states $\ket{(N-n)_g, n_e, 0_r}$
and $\ket{N_g, 0_e, 0_r}$. A related effect is described in \cite{Honer2011},
where the dephasing of collective Rydberg excitation was brought about by 
an inhomogeneous light field with short-range space and time correlations. 
Here, instead, the superposition containing a Rydberg atom is dephased 
by the field $\Om_{ge}$ and decay $\Gamma_{eg}$ through the cycling 
transition $\ket{g} \leftrightarrow \ket{e}$ of all the atoms that 
do not populate the Rydberg state. 

The field $\Om_{ge}(t)$ driving the $N-1$ blocked (two-level) atoms 
is spatially-uniform but varies slowly in time. At any time $t$, 
the average populations of state $\ket{g}$ for all the atoms  
is therefore approximately given by the steady-state expression 
for an independent two-level atom \cite{PLDP2006}, 
$\expv{\sig_{gg}} \approx 
\frac{\Om_{ge}^2 + \Ga_{eg}^2/4}{2 \Om_{ge}^2 + \Ga_{eg}^2/4} \equiv \kappa$.
The probability that all but the Rydberg excited atom are in 
the ground state $\ket{g}$ is then
$\expv{\sig_{rr}^j \prod_{i \neq j}^{N} \sig_{gg}^i} \approx [\kappa(t)]^{N-1}$.
At large times $\kappa(t_{\mathrm{end}}) \simeq 1$ and the above 
probability approaches unity, with the system in the mixed state
$\hrho = \frac{1}{N} \sum_{j=1}^N \sig_{rr}^j \prod_{i \neq j}^{N}\sig_{gg}^i$.
Yet, the total probability of finding one Rydberg excitation is 
$P_r(1) = \expv{\Sig_{r}^{(1)}} \simeq 1$ already when $\Om_{ge} > \Om_{er}$ 
[Fig.~\ref{fig:alsOmPr}]. These results are fully reproduced by the 
exact numerical solution of the density matrix equations~(\ref{rhoME}).

\subsection{Quantum Zeno effect}

An alternative and perhaps more elegant explanation as to why 
the superatom attains near unity Rydberg excitation 
$\expv{\Sig_r^{(1)}} \simeq 1$ is based on the quantum Zeno effect \cite{Zeno}:
Upon repetitive excitation to state $\ket{e}$ and spontaneous decay 
back to the ground state $\ket{g}$, the Rydberg blocked atoms perform 
continuous projective measurements of the Rydberg excitation in the ensemble. 

To verify this physical picture, we have performed quantum Monte Carlo 
simulations \cite{qjumps} of the dissipative dynamics of a few-atom system.
In such simulation, the state of the system $\ket{\Psi}$ evolves 
according to the Schr\"odinger equation 
$\partial_t \ket{\Psi} = -\frac{i}{\hbar} \tilde{\mc{H}} \ket{\Psi}$
with an effective non-Hermitian Hamiltonian 
$\tilde{\mc{H}} = \mc{H} - i \hbar \sum_j \hlf (\Ga_{eg} \sig_{ee}^j
+ \Ga_{re} \sig_{rr}^j)$ which does not preserve the norm of $\ket{\Psi}$. 
The evolution is interrupted by random quantum jumps 
$\ket{\Psi} \to \sig_{ge(er)}^j \ket{\Psi}$ with probabilities 
determined by the decay rates $\Ga_{eg(re)}$.

In the early part of evolution, the states of all the atoms share 
the same overlap with the current dark and bright states, and the state 
of the system $\ket{\Psi}$ is symmetric under permutation of the atoms. 
But already the first quantum jump breaks this symmetry by transferring 
one randomly selected atom from $\ket{e}$ to $\ket{g}$ (or, with a much 
smaller probability $\sim \Ga_{re}/\Ga_{eg}$, from $\ket{r}$ to $\ket{e}$), 
while the dark state overlap increases for the atoms that did not jump. 
During the subsequent evolution under the Rydberg blockade, the bright 
state overlap grows for all the atoms, while the following jump again 
suddenly increases the dark state contribution to the states of the atoms 
which did not jump. This proceeds untill eventually only one atom 
has experienced no quantum jump. This atom closely follows the dark 
state superposition, and while small excursions away from the dark 
state occur, they are reduced by each jump of the other atoms. 
Hence, atoms undergoing quantum jumps stabilize the almost 
deterministic evolution of a non-decaying atom towards the Rydberg state.

\subsection{Coherence relaxation}

\begin{figure}[t]
\includegraphics[width=7cm]{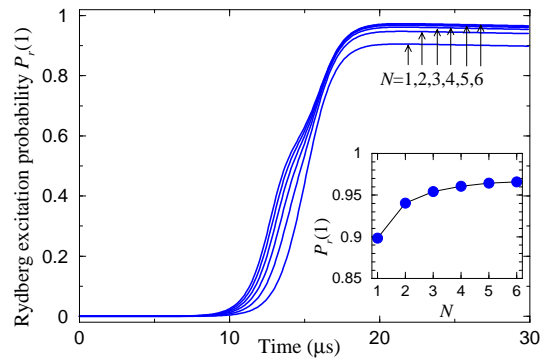}
\caption{
Probabilities $P_r(1)$ of single Rydberg excitation of superatom 
composed of $N = 1-6$ atoms, obtained from the exact solutions 
of Eq.~(\ref{rhoME}) including dephasing $\ga_{r}$ and decays 
$\Ga_{eg},\Ga_{re}$. Main panel shows the time-dependence of $P_r(1)$, 
and inset shows $P_r(1)$ for different $N$ at time 
$t_{\mathrm{end}} = 30 \:\mu$s.} 
\label{fig:Pr1garg}
\end{figure}

We finally demonstrate the robustness of adiabatic passage in a Rydberg 
superatom. In a single three-level atom, the STIRAP---while immune to 
the decay and moderate detuning of the intermediate state $\ket{e}$---is
very sensitive to coherence relaxation between the long-lived states 
$\ket{g}$ and $\ket{r}$ \cite{stirap}. 
The physical origins of relaxation of the atomic coherence include 
non-radiative collisions, Doppler shifts, laser phase fluctuations 
and electromagnetic field noise. 
In addition to the small decay $\Ga_{re}$ of the Rydberg state $\ket{r}$, 
we now include its dephasing $\ga_{r}$ via the Liouvillian 
$\mc{L}_{r}^j \hrho = \frac{1}{2}\ga_{r} 
[(\sig_{rr}^j - \sig_{ee}^j - \sig_{gg}^j) \hrho 
(\sig_{rr}^j - \sig_{ee}^j - \sig_{gg}^j) - \hrho]$ \cite{PLDP2006}.
In Fig.~\ref{fig:Pr1garg} we show the probability of Rydberg excitation of 
the superatom obtained from the solution of the master equation (\ref{rhoME})
with the dephasing rate $\ga_{r} = 2 \pi \times 0.1\:$MHz. 
For a single atom, the excitation probability is now reduced 
to $\expv{\sig_{rr}} \simeq 0.9$, since the decoherence of the dark 
state superposition leads to the population of the bright states \cite{EIT}. 
With increasing the number of atoms $N$, however, the Rydberg excitation
probability of the superatom grows according to $\expv{\Sig_r^{(1)}} 
\approx \frac{N \expv{\sig_{rr}}}{(N-1) \expv{\sig_{rr}} + 1}$
\cite{Petrosyan2012}, approaching $P_r(1) \gtrsim 0.97$ for $N =6$. 
The spontaneous decay, perhaps surprisingly, counteracts the detrimental 
effect of decoherence of the dark state superposition and facilitates 
the efficient production of a single Rydberg excitation. 
A similar result has been obtained for Rydberg superatoms composed 
of incoherently driven two-level atoms \cite{Petrosyan2012}.

\section{Discussion}

To conclude, we have examined the excitation of a Rydberg superatom
using adiabatic passage with delayed laser pulses. We have found 
that spontaneous decay of atoms from the intermediate excited 
state facilitates a single Rydberg excitation of the superatom, with 
nearly unit probability. An ensemble of $N > 1$ atoms in a tight trap 
of linear dimension smaller than the Rydberg blockade distance can 
repeatedly and reliably produce single Rydberg atoms or ions \cite{RdSV2013},
which can then be extracted and possibly deposited elsewhere. This 
process can proceed down to $N = 1$ atoms, and the final count of the 
extracted Rydberg atoms would correspond to the initial number of 
ground state atoms in the trap.

Experiments with the atomic ensembles much larger than the blockade 
length have revealed significant suppression of the number of Rydberg 
excitations \cite{Vogt2006,Tong2004,Singer2004,Heidemann2007}, which 
is consistent with a regular spatial arrangement of Rydberg atoms 
\cite{Schwarzkopf2011,Schauss2012}. We envisage that employing
stimulated adiabatic passage to produce with unit probability
singe Rydberg atoms per blockade volume can result in long-range 
order and tighter crystallization of Rydberg excitations in extended 
systems \cite{Petrosyan2012,Hoening2013} and may shine more light
onto the spatial correlation patterns of Rydberg excitations. 

Finally, our studies provide a new element to the very active field 
of dissipative generation of states and processes in open many-body systems
\cite{Diehl2008,Verstraete2009}. In our many-particle system, dissipation 
assigns different roles to different particles, such that some particles 
become dissipative probes for the coherent dynamics of the others. 
The observation that the decay of the probe particles may counteract 
decoherence in the target particle may find application in, e.g., 
ancilla assisted protocols for quantum computing \cite{Andersson}.

\begin{acknowledgments} 
K.M. gratefully acknowledges support from the FET-Open grant MALICIA (265522).
\end{acknowledgments}

\end{document}